# Identifying Reference Objects by Hierarchical Clustering in Java Environment

**Rahul Saha , Dr. G. Geetha**

**Department of Computer Science and Engineering, Lovely Professional University**

**Phagwara, Punjab, India**

*Email: rsahaaot@yahoo.com,*

**Department of Computer Sciences and Applications, Lovely Professional University**

**Phagwara, Punjab, India**

*Email: gitaskumar@yahoo.com*

## Abstract

Recently Java programming environment has become so popular. Java programming language is a language that is designed to be portable enough to be executed in wide range of computers ranging from cell phones to supercomputers. Computer programs written in Java are compiled into Java Byte code instructions that are suitable for execution by a Java Virtual Machine implementation. Java virtual Machine is commonly implemented in software by means of an interpreter for the Java Virtual Machine instruction set. As an object oriented language, Java utilizes the concept of objects. Our idea is to identify the candidate objects' references in a Java environment through hierarchical cluster analysis using reference stack and execution stack.

## 1. Introduction

Candidate Objects are those which can be selected as for the options for the objects in a object oriented paradigm. Object identification is a reverse-engineering technique that is largely used to assist the software migration from procedural paradigm to object-oriented paradigm. Object identification facilitates acquiring a precise knowledge of the data items in a program. Object identification reduces the degradation of original design. Object identification typically aims at finding match-up of legacy software components: data structures, and functions, for later building them as object-oriented classes. However, large application consists of numerous data structures and functions; it needs a statistical method to facilitate information classification.

## 2. Existing Concept

In the paper [1], the authors have described an approach of hierarchical clustering in a procedural language environment using stack and queues. Here the basic functions of a stack and queue are taken to create proximity matrix and pattern matrix. Pattern matrix represents a property set of data (scores or measurements) in a table. Each row stands for a set of properties (a pattern). Proximity matrix represents an index of association (proximity) between pair of patterns. The index can be either similarity index or dissimilarity index and can be computed. by several ways, for example, Simple matching coefficient, Jaccard coefficient, Euclidean distance, Manhattan distance etc. Then they have calculated Euclidean distances between each pair and least values are classified in a cluster. This process goes on until all the properties are successfully clustered. The extracted functions are shown in Table 1 and the relation definitions are shown in Table 2.

Table.1 : Extracted functions

| Software component |
|---|
| struct stack |
| struct queue |
| struct stack * initStack (int size) |
| struct queue *initQ ( ) |
| int isEmptyStack( struct stack * s) |
| int isEmptyQ ( struct queue * q) |
| void push ( struct stack * s, int i) |
| void enQ ( struct queue * q, int i) |
| int pop ( struct stack * s) |
| int deQ ( struct queue * q) |

Table.2: Relation Definition

| Name | Definition |
|---|---|
| R0 | Return type is struct stack |
| R1 | Return type is struct queue |
| R2 | Has argument of type struct stack |
| R3 | Has argument of type struct queue |
| R4 | Use field of struct stack |
| R5 | Use field of struct queue |

Now in Table 3 the extracted functions and the related definitions are shown. A ' X' mark is put in the cells for each corresponding function and relation definition. In Table 4 the cells are assigned by the values 0 or 1 depending upon the 'X' marks of Table 3; 1 is assigned for a 'X' mark else 0.

Table.3: Properties in modular case

|  | R0 | R1 | R2 | R3 | R4 | R5 |
|---|---|---|---|---|---|---|
| **initStack** | X |  |  |  | X |  |
| **initQ** |  | X |  |  |  | X |
| **isEmptyStack** |  |  | X |  | X |  |
| **Push** |  |  | X |  | X |  |
| **enQ** |  |  |  | X |  | X |
| **Pop** |  |  | X |  | X |  |
| **deQ** |  |  |  | X |  | X |

Table. 4: Pattern matrix for modular case

|  | R0 | R1 | R2 | R3 | R4 | R5 |
|---|---|---|---|---|---|---|
| **initStack** | 1 | 0 | 0 | 0 | 1 | 0 |
| **initQ** | 0 | 1 | 0 | 0 | 0 | 1 |
| **isEmptyStack** | 0 | 0 | 1 | 0 | 1 | 0 |
| **isEmptyQ** | 0 | 0 | 0 | 1 | 0 | 1 |
| **Push** | 0 | 0 | 1 | 0 | 1 | 0 |
| **enQ** | 0 | 0 | 0 | 1 | 0 | 1 |
| **Pop** | 0 | 0 | 1 | 0 | 1 | 0 |
| **deQ** | 0 | 0 | 0 | 1 | 0 | 1 |

Now a proximity matrix is generated using Euclidean-distance method formula given below:

$$d(i,k) = \left\{ \sum_{j=1}^{t} (x_{ij} - x_{kj})^2 \right\}^{1/2} \quad \ldots\ldots\ldots\ldots(\text{Equation 1})$$

where, $x_{ij}$ represents the j-ordered attribute of pattern i, $x_{kj}$

represents the j-ordered attribute of pattern k, and $t$ represents the total attribute of pattern. Depending on this proximity matrix clusters are made further by agglomerative method.

## 3. Our Approach

The authors have illustrated the above in case of procedural language. Our idea is to convert it into object orientation (Java environment).We shall use the same approach but using two stacks i.e. execution stack and reference stack so that the basic approach of the stacks as we have seen in the existing scenario above will be similar and we can integrate it with our Java environment. Table 5 and 6 show the relation definition and extracted functionalities used for our approach.

Table.5: Relation Definition of our approach

| Name | Definition |
|---|---|
| R0 | Return type is struct execstack |
| R1 | Return type is struct refstack |
| R2 | Has argument of type struct execstack |
| R3 | Has argument of type struct refstack |
| R4 | Use field of struct execstack |
| R5 | Use field of struct refstack |

Now, in Table 6 we have extracted some of the functionalities related to the stacks used in our approach that is the functionalities regarding reference stack and execution stack.

Table.6: Extracted functionalities for our approach

| Software component |
|---|
| struct execstack |
| struct refstack |
| struct execstack * initExec (int size ) |
| struct refstack *initRef ( int size ) |
| int isEmptyExec( struct execstack * es ) |
| int isEmptyRef ( struct refstack* rs) |
| void ePush ( struct execstack * es, int i ) |
| void rPush ( struct refstack * rs, int i ) |
| int ePop ( struct execstack * es ) |
| int rPop ( struct refstack * rs ) |
| struct execstack* traExec ( struct execstack * es ) |
| struct refstack * traRef (struct refstack* rs ) |

We can use a flowchart diagram to show that how reference and execution stacks are used in identifying the reference to objects. The diagram is given in Fig. 1.

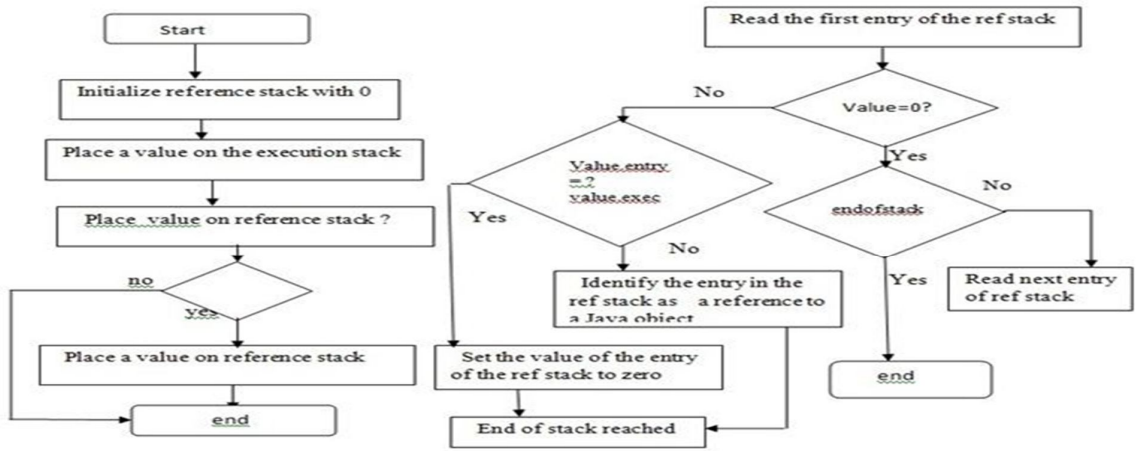

Fig. 1 Flowchart diagram of identifying reference objects

# 4. Analysis

The relation definition matrix is shown in table 7. We have also generated the pattern matrix which is shown in Table 8.

Table.7: Relation definition matrix for our approach

|            | R0 | R1 | R2 | R3 | R4 | R5 |
|------------|----|----|----|----|----|----|
| initRef    | X  |    |    |    | X  |    |
| initExec   |    | X  |    |    |    | X  |
| isEmptyRef |    |    |    | X  |    | X  |
| isEmptyExec|    |    | X  |    | X  |    |
| ePush      |    |    | X  |    | X  |    |
| rPush      |    |    |    | X  |    | X  |
| ePop       |    |    | X  |    | X  |    |
| rPop       |    |    |    | X  |    | X  |
| traRef     |    | X  |    | X  |    | X  |
| traExec    | X  |    | X  |    | X  |    |

Table.8: Pattern Matrix for our approach

|             | R0 | R1 | R2 | R3 | R4 | R5 |
|-------------|----|----|----|----|----|----|
| initRef     | 0  | 1  | 0  | 0  | 0  | 1  |
| initExec    | 1  | 0  | 0  | 0  | 1  | 0  |
| isEmptyRef  | 0  | 0  | 0  | 1  | 0  | 1  |
| isEmptyExec | 0  | 0  | 1  | 0  | 1  | 0  |
| ePush       | 0  | 0  | 1  | 0  | 1  | 0  |
| rPush       | 0  | 0  | 0  | 1  | 0  | 1  |
| ePop        | 0  | 0  | 1  | 0  | 1  | 0  |
| rPop        | 0  | 0  | 0  | 1  | 0  | 1  |
| traRef      | 0  | 1  | 0  | 1  | 0  | 1  |
| traExec     | 1  | 0  | 1  | 0  | 1  | 0  |

Now to generate the proximity matrices in each iteration we have used Euclidean formula as defined in Equation 1 earlier. The formula is as below once again:

$$d(i,k) = \left\{ \sum_{j=1}^{t} (x_{ij} - x_{kj})^2 \right\}^{1/2}$$

where, $x_{ij}$ represents the j-ordered attribute of pattern i, $x_{kj}$ represents the j-ordered attribute of pattern k, and $t$ represents the total attribute of pattern. First proximity matrix is shown in the Table 9.

Table.9: First proximity matrix

|             | initRef | initExec | isEmptyRef | isEmptyExec | epush | rpush | epop | rpop | traRef | traExec |
|-------------|---------|----------|------------|-------------|-------|-------|------|------|--------|---------|
| initRef     | 0       |          |            |             |       |       |      |      |        |         |
| initExec    | 2.00    | 0        |            |             |       |       |      |      |        |         |
| isEmptyRef  | 1.41    | 2.00     | 0          |             |       |       |      |      |        |         |
| isEmptyExec | 2.00    | 1.41     | 2.00       | 0           |       |       |      |      |        |         |
| ePush       | 1.41    | 1.41     | 2.00       | 0.00        | 0     |       |      |      |        |         |
| rPush       | 1.41    | 2.00     | 0.00       | 2.00        | 2.00  | 0     |      |      |        |         |
| ePop        | 2.00    | 1.41     | 2.00       | 0.00        | 0.00  | 2.00  | 0    |      |        |         |
| rPop        | 1.41    | 2.00     | 0.00       | 2.00        | 2.00  | 0.00  | 2.00 | 0    |        |         |
| traRef      | 1.00    | 2.24     | 2.00       | 2.24        | 2.24  | 1.00  | 2.24 | 1.00 | 0      |         |
| traExec     | 2.24    | 1.00     | 2.24       | 1.00        | 1.00  | 2.24  | 1.00 | 2.24 | 2.45   | 0       |

We have considered the least distant (0.00) values from the Table 9 first to form the first round clusters and applied Single linkage rule to form the second proximity matrix in Table 10. The formula for the single linkage rule goes thus:

$d [(k), (i,j)] = min \{d [(k),(i)], d [(k),(j)] \}$

where:

d [(k),(i)] represents the similarity between cluster k and cluster i d [(k),j )] represents the similarity between cluster k and cluster j d [(k),(i, j)] represents the similarity between cluster k and the newly formed cluster i, j.

Table.10: Second Proximity Matrix

|  | initRef | initExec | C1 | C2 | traRef | traExec |
|---|---|---|---|---|---|---|
| initRef | 0 | | | | | |
| initExec | 2.00 | 0 | | | | |
| C1 | 1.41 | 2.00 | 0 | | | |
| C2 | 1.41 | 1.41 | 2.00 | 0 | | |
| traRef | 1.00 | 2.24 | 1.00 | 2.24 | 0 | |
| traExec | 2.24 | 1.00 | 2.24 | 1.00 | 2.45 | 0 |

C1 = isEmptyRef + rPush + rPop
C2 = isEmptyExec + ePush + ePop

We now have considered the least distant (1.00) values from the Table 10 first to form the next round of clusters and applied the above said Single linkage rule to form the third proximity matrix given in Table 11.

Table.11: Third Proximity Matrix

|  | C1 | C2 | C3 | C4 |
|---|---|---|---|---|
| C1 | 0 | | | |
| C2 | 2.00 | 0 | | |
| C3 | 1.00 | 1.41 | 0 | |
| C4 | 2.00 | 1.00 | 2.00 | 0 |

C3 = traRef + initRef   C4 = traExec + initExec

We now have considered the least distant (1.00) values from the Table 11 to form the next round of clusters and applied Single linkage rule to form the fourth proximity matrix given in Table 12.

Table. 12 Fourth Proximity Matrix

|  | C5 | C6 |
|---|---|---|
| C5 | 0 | |
| C6 | 1.41 | 0 |

C5=C1 + C3     C6=C2+ C4

We now have considered the least distant (0.00) values from the Table 12 to form the next round of clusters and applied Single linkage rule to form the fifth proximity matrix shown in Table 13.

Table.13: Fifth Proximity Matrix

|  | C7 |
|---|---|
| C7 | 0 |

C7 = C5 + C6

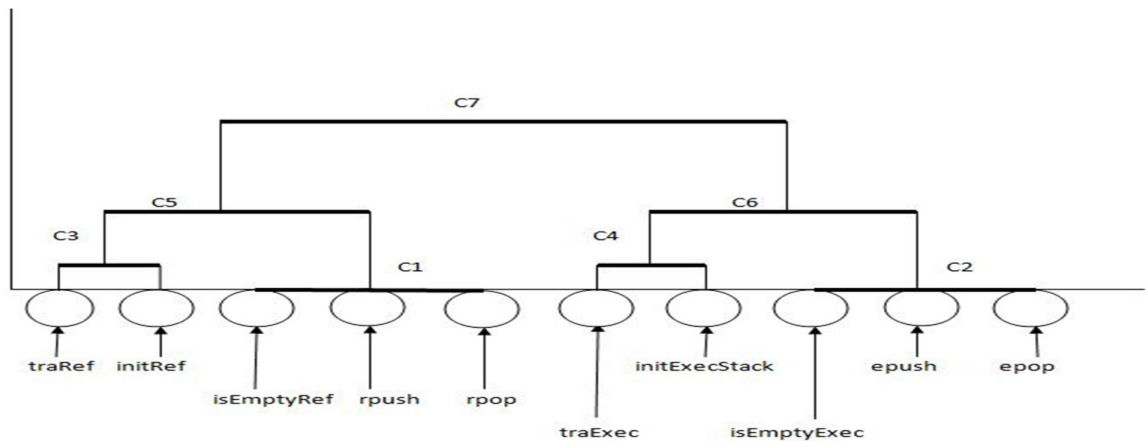

Fig. 2 Dendogram of clusters

## 4.1. Dendogram

Output of clustering analysis can be represented in various forms depending on the objective of data classification. Output of hierarchical clustering analysis is usually represented in a special type of a tree structure called Dendogram. Our output of clusterization is also represented in Dendogram shown in Fig. 2 above.

## 5. Conclusion

From the above Dendogram, we can see that the cluster C5 contains all the functionalities that deal with reference stack and cluster C6 contains all the functionalities that deal with execution stack. Cluster C5 can be further divided into C3 and C1 where C3 cluster consists of the functionalities of traversing the reference stack i.e. traRef ( ) and initialization of reference stack i.e. initRef ( ). Cluster C1 consists of the functionalities like to check if the reference stack is empty [ isEmptyRef ( ) ], to insert objects in the reference stack [ rpush ( ) ] , and to delete an object reference [ rpop( ) ].

Similarly, cluster C6 can be further divided into cluster C4 and cluster C2 where C4 consists of traExec ( ) [ traversing the execution stack ], initExec ( ) [ initialization of execution stack ] and cluster C2 consists of isEmptyExec ( ), epush( ) and epop( ).

## Authors' profile

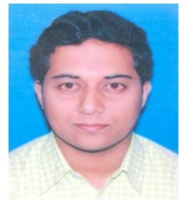

**Rahul Saha** is pursuing his M.Tech from Lovely Professional University, Punjab, India in the department of Computer Science and Engineering. His research interest includes Software Engineering, Network Security.

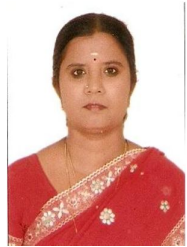

**Dr. G. Geetha** is the Dean of School of Computer Sciences and Applications. Her research interest includes Cryptography and Software Engineering. She has published more than 30 papers in refereed Journals and Conferences. She is also the Editorial Board of IJACM and IJCRYPTO. She is presently the President of Advanced Computing Research Society.